\documentclass[12pt]{article}
\usepackage{amsmath,amsfonts,amssymb}

\begin{document}
\thispagestyle{empty}

\def\theequation{\arabic{section}.\arabic{equation}}
\def\a{\alpha}
\def\b{\beta}
\def\g{\gamma}
\def\d{\delta}
\def\dd{\rm d}
\def\e{\epsilon}
\def\ve{\varepsilon}
\def\z{\zeta}
\def\B{\mbox{\bf B}}

\newcommand{\h}{\hspace{0.5cm}}

\begin{titlepage}
\vspace*{1.cm}
\renewcommand{\thefootnote}{\fnsymbol{footnote}}
\begin{center}
{\Large \bf Three-point correlators: finite-size giant magnons and
singlet scalar operators on higher string levels}
\end{center}
\vskip 1.2cm \centerline{\bf Plamen Bozhilov} \vskip 0.6cm
\centerline{\sl Institute for Nuclear Research and Nuclear Energy}
\centerline{\sl Bulgarian Academy of Sciences} \centerline{\sl  1784
Sofia, Bulgaria}

\centerline{\tt plbozhilov@gmail.com}

\vskip 20mm

\baselineskip 18pt

\begin{center}
{\bf Abstract}
\end{center}
\h In the framework of the semiclassical approach, we compute the
normalized structure constants in three-point correlation functions,
when two of the vertex operators correspond to "heavy" string
states, while the third vertex corresponds to a "light" state. This
is done for the case when the "heavy" string states are {\it
finite-size} giant magnons, carrying one or two angular momenta. The
"light" states are taken to be singlet scalar operators on higher
string levels. We first consider the case of string theory on
$AdS_5\times S^5$ dual to $\mathcal{N} = 4$ super Yang-Mills. Then
we extend the obtained results to the $\gamma$-deformed $AdS_5\times
S^5_\gamma$, corresponding to $\mathcal{N} = 1$ super Yang-Mills
theory, appearing as an exactly marginal deformation of $\mathcal{N}
= 4$ super Yang-Mills.

\end{titlepage}
\newpage

\def\nn{\nonumber}
\def\tr{{\rm tr}\,}
\def\p{\partial}
\newcommand{\bea}{\begin{eqnarray}}
\newcommand{\eea}{\end{eqnarray}}
\newcommand{\bde}{{\bf e}}
\renewcommand{\thefootnote}{\fnsymbol{footnote}}
\newcommand{\be}{\begin{equation}}
\newcommand{\ee}{\end{equation}}

\vskip 0cm

\renewcommand{\thefootnote}{\arabic{footnote}}
\setcounter{footnote}{0}


\setcounter{equation}{0}
\section{Introduction}

There have been many investigations on correlation functions in the
AdS/CFT context \cite{AdS/CFT}. Recently, several interesting
developments have been made by considering general "heavy" string
states. An efficient method to compute two-point correlation
functions in the strong coupling limit is to evaluate string
partition function for a "heavy" string state propagating in the AdS
space between two boundary points based on a path integral approach
\cite{Janik,BuchTsey}. This method has been extended to the
three-point functions of two "heavy" string states and a "light"
mode \cite{Zarembo,Costa,rt10}. Relying on these achievements, many
interesting results concerning correlators of "heavy" and "light"
modes have been obtained \cite{Zarembo}-\cite{MRS1107}.

As explained in \cite{rt10}, there exist special massive string
states vertex operators with finite  quantum numbers  for which the
leading-order bosonic part is known explicitly and thus they can be
used as candidates for ``light'' vertex operators in the
semiclassical computation of the correlation functions. These are
singlet operators which do not mix  with other operators to leading
nontrivial order in $\frac{1}{\sqrt{\lambda}}$  \cite{tsev,rt9}. An
example of such scalar operator carrying no spins is
\cite{rt10}\footnote{ The marginality condition for  this operator
is \cite{rt10}:
 $2(1-q)   +  \frac{1}{2\sqrt{\lambda}}\Big[\Delta (\Delta-4) +  2q(q-1)\Big]
\\ \nn  +  {\frac{1}{(\sqrt{\lambda})^2} } \big[ \frac{2}{3} q (q-1) (q-
\frac{7}{2})  +
 4q \big]  +\mathcal{O}\left(\frac{1}{(\sqrt{\lambda})^3}\right)=0$.}
 \bea\label{Vq}   V_q= (Y_4+Y_5)^{- \Delta} \Big[(\p X_k \bar{\p} X_k)^q+ ... \Big]
.\eea This  operator  corresponds to a scalar string state at level
$n=q-1$ so that the fermionic contributions   should make the  $q=1$
state massless (BPS), with $\Delta=4$  following from the
marginality condition. The $q=2$ choice   corresponds to a scalar
state on the first excited string level. In that case, we have
\cite{rt9} \bea\nn
\Delta(\Delta-4)=4(\sqrt{\lambda}-1)+\mathcal{O}\left(\frac{1}{\sqrt{\lambda}}\right),\eea
with solution \bea\nn
\Delta=2(\lambda^{1/4}+1)+\frac{0}{\lambda^{1/4}}+\mathcal{O}\left(\frac{1}{\lambda^{3/4}}\right).\eea
However, the subleading terms here should not be trusted as far as
the fermions are expected to change the $\Delta$-independent terms
in the 1-loop anomalous dimension. For arbitrary string level $n$,
the solution of the marginality condition with respect to $\Delta$,
to leading order in $\frac{1}{\sqrt{\lambda}}$, is given by
\bea\label{mc}
\Delta_q=2\left(\sqrt{(q-1)\sqrt{\lambda}+1-\frac{1}{2}q(q-1)}+1\right).\eea

Let us also point out that the number $q$ of $\p X_k \bar{\p} X_k$
factors in an operator never increases due to
renormalization~\cite{tsev}. That is why, it can be used as a
quantum number to characterize the leading term in the corresponding
operator \cite{Wegner90}.

Here we will be interested in semiclassical computation of
three-point correlation functions for the case when the "heavy"
string states are {\it finite-size} giant magnons, carrying one or
two angular momenta, while the "light" states are taken to be given
by (\ref{Vq}) for $q=1,2,3,...$. First, we will consider the case of
giant magnons in $AdS_5\times S^5$. Then we will extend the obtained
results to giant magnons on $\gamma$-deformed $AdS_5\times
S^5_\gamma$ background. To this end, we will use the following
approach \cite{rt10,Hernandez2}. The three-point functions of two
"heavy" operators and a "light" operator can be approximated by a
supergravity vertex operator evaluated at the "heavy" classical
string configuration: \bea \nn \langle
V_{H}(x_1)V_{H}(x_2)V_{L}(x_3)\rangle=V_L(x_3)_{\rm classical}. \eea
For $\vert x_1\vert=\vert x_2\vert=1$, $x_3=0$, the correlation
function reduces to \bea \nn \langle
V_{H_1}(x_1)V_{H_2}(x_2)V_{L}(0)\rangle=\frac{C_{123}}{\vert
x_1-x_2\vert^{2\Delta_{H}}}. \eea Then, the normalized structure
constants \bea \nn \mathcal{C}_3=\frac{C_{123}}{C_{12}} \eea can be
found from \bea \label{nsc} \mathcal{C}_3=c_{\Delta}V_L(0)_{\rm
classical}, \eea were $c_{\Delta}$ is the normalized constant of the
corresponding "light" vertex operator. Actually, we are going to
compute the corresponding normalized structure constants
(\ref{nsc}).

\setcounter{equation}{0}
\section{Three-point correlators for giant magnons on \\ $AdS_5\times S^5$}

Let us first introduce the coordinates, which we are going to use
further on. If we denote the string embedding coordinates on $AdS$
and $S^5$ parts of the $AdS_5\times S^5$ background with $Y$ and $X$
respectively, then \bea\nn &&Y_1+iY_2=\sinh\rho\ \sin\eta\
e^{i\varphi_1},\h Y_3+iY_4=\sinh\rho\ \cos\eta\ e^{i\varphi_2},\h
Y_5+iY_0=\cosh\rho\ e^{it}, \eea are related to the Poincare
coordinates by
 \bea \nn Y_m=\frac{x_m}{z},\h
Y_4=\frac{1}{2z}\left(x^mx_m+z^2-1\right), \h
Y_5=\frac{1}{2z}\left(x^mx_m+z^2+1\right), \eea where $x^m
x_m=-x_0^2+x_ix_i$, with $m=0,1,2,3$ and $i=1,2,3$.

For giant magnons, the $AdS$ part of the solution, after Euclidean
rotation, is given by \cite{Costa,Hernandez2} ($i\tau=\tau_e$, where
$\tau$ is the worldsheet time and $\kappa$ is a parameter) \bea\nn
&&x_{0e}=\tanh(\kappa\tau_e),\h x_i=0,\h
z=\frac{1}{\cosh(\kappa\tau_e)}.\eea Thus, the factor $(Y_4+Y_5)^{-
\Delta}$ in (\ref{Vq}) becomes
$\left(\cosh(\kappa\tau_e)\right)^{-\Delta}$, while $\p X_k \bar{\p}
X_k$ is basically the string Lagrangian on the five-sphere, computed
on the giant magnons' first integrals.

Since we are going to find the three-point correlators containing
two heavy operators corresponding to giant magnons with one or two
angular momenta, we restrict ourselves to the three-sphere. Then, we
can explore the reduction of the string dynamics to the
Neumann-Rosochatius integrable model by using the ansatz
\cite{KRT06} \bea\label{NRA} &&t(\tau,\sigma)=\kappa\tau,\h
\theta(\tau,\sigma)=\theta(\xi),\h
\phi_j(\tau,\sigma)=\omega_j\tau+f_j(\xi),\\ \nn
&&\xi=\alpha\sigma+\beta\tau,\h \kappa, \omega_j, \alpha,
\beta=constants,\h j=1,2.\eea As a consequence, the string
Lagrangian in conformal gauge, on the three-sphere, can be written
as (prime is used for $d/d\xi$) \bea\nn
&&\mathcal{L}_{S^3}=(\alpha^2-\beta^2)
\left[\theta'^2+\sin^2\theta\left(f'_1-\frac{\beta\omega_1}{\alpha^2-\beta^2}\right)^2
+\cos^2\theta\left(f'_2-\frac{\beta\omega_2}{\alpha^2-\beta^2}\right)^2
\right.
\\ \label{rl} &&-\left.\frac{\alpha^2}{(\alpha^2-\beta^2)^2}
\left(\omega_1^2\sin^2\theta+\omega_2^2\cos^2\theta\right)
\right].\eea One can prove that the first integrals of the equations
of motion for $f_j(\xi)$, $\theta(\xi)$, take the form \bea\nn
&&f'_1=\frac{\omega_1}{\alpha}\frac{v}{1-v^2} \left(\frac{
W}{1-\chi} -1\right),
\\ \label{tfi} &&f'_2=-\frac{\omega_1}{\alpha}\frac{uv}{1-v^2} ,
\\ \nn && \theta'
=\frac{\omega_1}{\alpha}\frac{\sqrt{1-u^{2}}}{1-v^2}
\sqrt{\frac{(\chi_{p}-\chi)(\chi-\chi_{m})}{1-\chi}},\eea where
\bea\nn &&\chi_p+\chi_m=\frac{2-(1+v^2)W-u^2}{1
-u^2},\\
\label{3eqs} &&\chi_p \chi_m=\frac{(1-W)(1-v^2W)}{1 -u^2}.\eea The
case of {\it finite-size} dyonic giant magnons, corresponds to
\bea\nn 0<u<1,\h 0<v<1,\h 0<W<1,\h 0<\chi_{m}<\chi< \chi_{p}<1.\eea
In (\ref{tfi}) and (\ref{3eqs}) the following notations have been
introduced \bea\nn &&\chi=\cos^2\theta, \h
v=-\frac{\beta}{\alpha},\h u=\frac{\omega_2}{\omega_1},\h
W=\left(\frac{\kappa}{\omega_1}\right)^2 .\eea The replacement into
(\ref{rl}) gives (we set $\alpha=\omega_1=1$ for simplicity)
\bea\label{Lgm} &&\mathcal{L}_{S^3}^{gm}=
-\frac{1}{1-v^2}\left[2-(1+v^2)W
-2\left(1-u^2\right)\chi\right].\eea We will need also the first
integral for $\chi$, which is given by \bea\label{fichi} \chi'=
\frac{2\sqrt{1-u^{2}}}{1-v^2}
\sqrt{\chi(\chi_{p}-\chi)(\chi-\chi_{m})}.\eea

After all that, the vertex (\ref{Vq}) can be rewritten as
\bea\label{Vqgm}   V_q=
\left(\cosh(\sqrt{W}\tau_e)\right)^{-\Delta_q}
\left(\mathcal{L}_{S^3}^{gm}\right)^{q} ,\eea where $\Delta_q$
should be taken from (\ref{mc}) and $\mathcal{L}_{S^3}^{gm}$ - from
(\ref{Lgm}). Then, according to \cite{rt10,Hernandez2}, the
normalized structure constant (\ref{nsc}) becomes \bea\label{c3d}
\mathcal{C}_3^q=c_{\Delta_q}\int_{-\infty}^{\infty}\frac{d\tau_e}{\cosh^{\Delta_q}(\sqrt{W}\tau_e)}
\int_{-L}^{L}d\sigma\left(\mathcal{L}_{S^3}^{gm}\right)^{q}.\eea
Here, the parameter $L$ is introduced to take into account the {\it
finite-size} of the giant magnons. Otherwise, $L\to\infty$.

The integration over $\tau_e$ in (\ref{c3d}) gives \bea\nn
\int_{-\infty}^{\infty}\frac{d\tau_e}{\cosh^{\Delta_q}(\sqrt{W}\tau_e)}=
\sqrt{\frac{\pi}{W}}\
\frac{\Gamma\left(\frac{\Delta_q}{2}\right)}{\Gamma\left(\frac{\Delta_q+1}{2}\right)},\eea
where $\Gamma(x)$ is the Euler gamma-function.
The integration over
$\sigma$ can be replaced by integration over $\chi$ according to
\bea\label{int} \int_{-L}^{L}d\sigma=
2\int_{\chi_m}^{\chi_p}\frac{d\chi}{\chi'},\eea where $\chi'$ is
given in (\ref{fichi}). Then (\ref{c3d}) acquires its final
form\bea\label{c3df} &&\mathcal{C}_3^q=c_{\Delta_q}\pi^{3/2}
\frac{\Gamma\left(\frac{\Delta_q}{2}\right)}{\Gamma\left(\frac{\Delta_q+1}{2}\right)}
\frac{(-1)^q\left[2-(1+v^2)W\right]^q}{(1-v^2)^{q-1}\sqrt{(1-u^2)W\chi_p}}
\\ \nn &&\sum_{k=0}^{q}\frac{q!}{k!(q-k)!}\left[-\frac{1-u^2}{1-\frac{1}{2}
(1+v^2)W}\right]^{k}\chi_p^{k}\
{}_2F_1\left(\frac{1}{2},\frac{1}{2}-k;1;1-\frac{\chi_m}{\chi_p}\right),\eea
where in accordance with (\ref{3eqs})  \bea\nn
&&\chi_p=\frac{1}{2(1-u^2)} \left\{q_1+q_2-u^2
+\sqrt{(q_1-q_2)^2-\left[2\left(q_1+q_2-2q_1 q_2\right)-u^2\right]
u^2}\right\},
\\ \nn &&\chi_m=\frac{1}{2(1-u^2)} \left\{q_1+q_2-u^2
-\sqrt{(q_1-q_2)^2-\left[2\left(q_1+q_2-2q_1 q_2\right)-u^2\right]
u^2}\right\},
\\ \label{roots} &&q_1=1-W,\h q_2=1-v^2W ,\eea
and ${}_2F_1\left(a,b;c;z\right)$ is the Gauss hypergeometric
function.

This is our general result corresponding to finite-size giant
magnons with two angular momenta and to arbitrary string level
$n=q-1=0,1,2,...$ . Now, let us give some particular examples
contained in (\ref{c3df}).

\subsection{Giant magnons with one angular momentum}
The case of finite-size giant magnons with one angular momentum
$J_1\ne 0$ corresponds to $u=0$. This can be seen from the explicit
expression for the second angular momentum $J_2$: \bea\nn
\mathcal{J}_2\equiv\frac{2\pi J_2}{\sqrt{\lambda}} =\frac{2u
\sqrt{\chi_{p}}}{\sqrt{1-u^2}}\
\mathbf{E}\left(1-\frac{\chi_m}{\chi_p}\right),\eea where
$\mathbf{E}(x)$ is the complete elliptic integral of second kind.
Then from (\ref{roots}) one obtains the following simplified
expressions for $\chi_p$, $\chi_m$: \bea\nn \chi_p=1-v^2W,\h
\chi_m=1-W.\eea Taking this into account, and using (\ref{c3df}),
one can find that the normalized structure constants for the first
three string levels, for the case at hand, are given by:

$q=1$ (level $n=0$) \bea\nn &&\mathcal{C}_3^1=2c_{\Delta_1}\pi^{1/2}
\frac{\Gamma\left(\frac{\Delta_1}{2}\right)}{\Gamma\left(\frac{\Delta_1+1}{2}\right)}
\frac{1}{\sqrt{W(1-v^2W)}}
\\ \nn &&\left[2(1-v^2W) \
\mathbf{E}\left(1-\frac{1-W}{1-v^2W}\right)
-\left(2-(1+v^2)W\right)\mathbf{K}\left(1-\frac{1-W}{1-v^2W}\right)\right].\eea
$q=2$ (level $n=1$) \bea\nn &&\mathcal{C}_3^2=2c_{\Delta_2}\pi^{1/2}
\frac{\Gamma\left(\frac{\Delta_2}{2}\right)}{\Gamma\left(\frac{\Delta_2+1}{2}\right)}
\frac{1}{(1-v^2)\sqrt{W(1-v^2W)}}
\\ \nn
&&\left[\left(2-(1+v^2)W\right)^2\mathbf{K}\left(1-\frac{1-W}{1-v^2W}\right)\right.
\\ \nn &&-4\left.\left(2-(1+v^2)W\right)(1-v^2W) \
\mathbf{E}\left(1-\frac{1-W}{1-v^2W}\right)\right.
\\ \nn &&+\left. 2\pi(1-v^2W)^2 \ {}_2F_1\left(\frac{1}{2},-\frac{3}{2};1;1-\frac{1-W}{1-v^2W}\right)\right].\eea
$q=3$ (level $n=2$) \bea\nn
&&\mathcal{C}_3^3=-2c_{\Delta_3}\pi^{1/2}
\frac{\Gamma\left(\frac{\Delta_3}{2}\right)}{\Gamma\left(\frac{\Delta_3+1}{2}\right)}
\frac{\left(2-(1+v^2)W\right)^3}{(1-v^2)^2\sqrt{W(1-v^2W)}}
\\ \nn &&\left[\mathbf{K}\left(1-\frac{1-W}{1-v^2W}\right)
-\frac{6(1-v^2W)}{2-(1+v^2)W} \
\mathbf{E}\left(1-\frac{1-W}{1-v^2W}\right)\right.
\\ \nn &&+\left.\frac{6\pi(1-v^2W)^2}{\left(2-(1+v^2)W\right)^2} \
{}_2F_1\left(\frac{1}{2},-\frac{3}{2};1;1-\frac{1-W}{1-v^2W}\right)\right.
\\ \nn &&-\left.\frac{4\pi(1-v^2W)^3}{\left(2-(1+v^2)W\right)^3} \
{}_2F_1\left(\frac{1}{2},-\frac{5}{2};1;1-\frac{1-W}{1-v^2W}\right)
\right].\eea $\mathbf{K}(x)$ in the expressions above is the
complete elliptic integral of first kind.

\subsection{Giant magnons with two angular momenta}

$q=1$ (level $n=0$): \bea\nn
&&\mathcal{C}_3^1=2c_{\Delta_1}\pi^{1/2}
\frac{\Gamma\left(\frac{\Delta_1}{2}\right)}{\Gamma\left(\frac{\Delta_1+1}{2}\right)}
\frac{1}{\sqrt{(1-u^2)W\chi_p}}
\\ \nn &&\left[2(1-u^2)\chi_p \
\mathbf{E}\left(1-\frac{\chi_m}{\chi_p}\right)
-\left(2-(1+v^2)W\right)\mathbf{K}\left(1-\frac{\chi_m}{\chi_p}\right)\right].\eea
$q=2$ (level $n=1$): \bea\nn
&&\mathcal{C}_3^2=2c_{\Delta_2}\pi^{1/2}
\frac{\Gamma\left(\frac{\Delta_2}{2}\right)}{\Gamma\left(\frac{\Delta_2+1}{2}\right)}
\frac{1}{(1-v^2)\sqrt{(1-u^2)W\chi_p}}
\\ \nn &&\left[\left(2-(1+v^2)W\right)^2\mathbf{K}\left(1-\frac{\chi_m}{\chi_p}\right)
-4(1-u^2)\left(2-(1+v^2)W\right)\chi_p \
\mathbf{E}\left(1-\frac{\chi_m}{\chi_p}\right)\right.
\\ \nn &&+\left. 2\pi(1-u^2)^2\chi_p^2 \ {}_2F_1\left(\frac{1}{2},-\frac{3}{2};1;1-\frac{\chi_m}{\chi_p}\right)\right].\eea
$q=3$ (level $n=2$): \bea\nn
&&\mathcal{C}_3^3=-2c_{\Delta_3}\pi^{1/2}
\frac{\Gamma\left(\frac{\Delta_3}{2}\right)}{\Gamma\left(\frac{\Delta_3+1}{2}\right)}
\frac{\left(2-(1+v^2)W\right)^3}{(1-v^2)^2\sqrt{(1-u^2)W\chi_p}}
\\ \nn &&\left[\mathbf{K}\left(1-\frac{\chi_m}{\chi_p}\right)
-\frac{6(1-u^2)\chi_p}{2-(1+v^2)W} \
\mathbf{E}\left(1-\frac{\chi_m}{\chi_p}\right)\right.
\\ \nn &&+\left.\frac{6\pi(1-u^2)^2\chi_p^2}{\left(2-(1+v^2)W\right)^2} \
{}_2F_1\left(\frac{1}{2},-\frac{3}{2};1;1-\frac{\chi_m}{\chi_p}\right)\right.
\\ \nn &&-\left.\frac{4\pi(1-u^2)^3\chi_p^3}{\left(2-(1+v^2)W\right)^3} \
{}_2F_1\left(\frac{1}{2},-\frac{5}{2};1;1-\frac{\chi_m}{\chi_p}\right)
\right].\eea

\setcounter{equation}{0}
\section{Three-point correlators for giant magnons on \\ $AdS_5\times S^5_{\gamma}$}

Investigations on AdS/CFT duality \cite{AdS/CFT} for the cases with
reduced or without supersymmetry is of obvious interest and
importance. An interesting example of such correspondence between
gauge and string theory models with reduced supersymmetry is
provided by an exactly marginal deformation of $\mathcal{N} = 4$
super Yang-Mills theory \cite{LS95} and string theory on a
$\beta$-deformed $AdS_5\times S^5$ background suggested by Lunin and
Maldacena in \cite{LM05}. When $\beta\equiv\gamma$ is real, the
deformed background can be obtained from $AdS_5\times S^5$ by the
so-called TsT transformation. It includes T-duality on one angle
variable, a shift of another isometry variable, then a second
T-duality on the first angle \cite{F05}. Taking into account that
the five-sphere has three isometric coordinates, one can consider
generalization of the above procedure, consisting of chain of three
TsT transformations. The result is a regular three-parameter
deformation of $AdS_5\times S^5$ string background, dual to a
non-supersymmetric deformation of $\mathcal{N} = 4$ super Yang-Mills
\cite{F05}, which is conformal in the planar limit to any order of
perturbation theory \cite{AKS06}. The action for this
$\gamma_i$-deformed $(i=1,2,3)$ gauge theory can be obtained from
the initial one after replacement of the usual product with
associative $*$-product \cite{LM05,F05,BR05}.

An essential property of the TsT transformation is that it preserves
the classical integrability of string theory on $AdS_5\times S^5$
\cite{F05}. The $\gamma$-dependence enters only through the {\it
twisted} boundary conditions and the {\it level-matching} condition.
The last one is modified since a closed string in the deformed
background corresponds to an open string on $AdS_5\times S^5$ in
general.

The bosonic part of the Green-Schwarz action for strings on the
$\gamma_i$-deformed $AdS_5\times S_{\gamma_i}^5$ \cite{AAF05}
reduced to $R_t\times S_{\gamma_i}^5$ can be written as (the common
radius $R$ of $AdS_5$ and $S_{\gamma_i}^5$ is set to 1)
\bea\label{BGS} S&=&-\frac{T}{2}\int d\tau
d\sigma\left\{\sqrt{-\gamma}\gamma^{ab}\left[-\p_a t\p_b t+\p_a
r_i\p_br_i+Gr_i^2\p_a\phi_i\p_b\phi_i \right. \right.\\ \nn &&
+\left.\left.Gr_1^2r_2^2r_3^2
\left(\tilde{\gamma}_i\p_a\phi_i\right)
\left(\tilde{\gamma}_j\p_b\phi_j\right) \right]\right.\\ \nn
&&-2G\left.\epsilon^{ab}\left(\tilde{\gamma}_3r_1^2r_2^2\p_a\phi_1\p_b\phi_2
+\tilde{\gamma}_1r_2^2r_3^2\p_a\phi_2\p_b\phi_3
+\tilde{\gamma}_2r_3^2r_1^2\p_a\phi_3\p_b\phi_1\right)\right\} ,\eea
where $T$ is the string tension, $\gamma^{ab}$ is the worldsheet
metric, $\phi_i$  are the three isometry angles of the deformed
$S_{\gamma_i}^5$, and \bea\label{roG} \sum_{i=1}^{3}r_i^2=1,\h
G^{-1}=1+\tilde{\gamma}_3^2r_1^2r_2^2+\tilde{\gamma}_1^2r_2^2r_3^2
+\tilde{\gamma}_2^2r_1^2r_3^2.\eea The deformation parameters
$\tilde{\gamma}_i$ are related to $\gamma_i$ which appear in the
dual gauge theory as follows \bea\nn \tilde{\gamma}_i = 2\pi T
\gamma_i = \sqrt{\lambda} \gamma_i .\eea When
$\tilde{\gamma}_i=\tilde{\gamma}$ this becomes the supersymmetric
background of \cite{LM05}, and the deformation parameter $\gamma$
enters the $\mathcal{N}=1$ SYM superpotential in the following way
\bea\nn W\propto
tr\left(e^{i\pi\gamma}\Phi_1\Phi_2\Phi_3-e^{-i\pi\gamma}\Phi_1\Phi_3\Phi_2\right).\eea
This is the case we are going to consider here.

Since the giant magnons with one or two angular momenta live on the
subspace $R_t\times S_{\gamma}^3$ $(r_3=0,\phi_3=0)$, we restrict
ourselves to that subspace of $AdS_5\times S_{\gamma}^5$ ,
parameterize (see (\ref{roG})) \bea\nn r_1=\sin\theta , \h
r_2=\cos\theta ,\eea and use the ansatz (\ref{NRA}). Then the string
Lagrangian in conformal gauge, on the $\gamma$-deformed
three-sphere, can be written as (prime is used for $d/d\xi$) \bea\nn
&&\mathcal{L}=(\alpha^2-\beta^2)
\left[\theta'^2+G\sin^2\theta\left(f'_1-\frac{\beta\omega_1}{\alpha^2-\beta^2}\right)^2
+G\cos^2\theta\left(f'_2-\frac{\beta\omega_2}{\alpha^2-\beta^2}\right)^2
\right.
\\ \label{r3} &&-\left.\frac{\alpha^2}{(\alpha^2-\beta^2)^2}G
\left(\omega_1^2\sin^2\theta+\omega_2^2\cos^2\theta\right)
+2\alpha\tilde{\gamma}G\sin^2 \theta \cos^2 \theta \frac{\omega_2
f'_1-\omega_1 f'_2}{\alpha^2-\beta^2}\right], \eea where \bea\nn
G=\frac{1}{1+\tilde{\gamma}^2\sin^2 \theta \cos^2 \theta}.\eea

The equations of motion for $f_{1,2}$ following from (\ref{r3}) can
be integrated once to give \bea\label{fjs}  &&
f'_1=\frac{1}{\alpha^2-\beta^2} \left[\frac{C_1}{\sin^2 \theta}
+\beta\omega_1-\tilde{\gamma}\left(\alpha\omega_2-\tilde{\gamma}C_1\right)\cos^2\theta\right],
\\ \nn &&f'_2=\frac{1}{\alpha^2-\beta^2} \left[\frac{C_2}{\cos^2 \theta}
+\beta\omega_2+\tilde{\gamma}\left(\alpha\omega_1+\tilde{\gamma}C_2\right)\sin^2\theta\right]
,\eea where $C_{1,2}$ are integration constants.

Next, we should take into account the Virasoro constraints, which
for our case are given by \bea\nn && \theta'^2+
G\sin^2\theta\left(f'^2_1+\frac{2\beta\omega_1}{\alpha^2+\beta^2}f'_1+\frac{\omega_1^2}{\alpha^2+\beta^2}\right)
\\ \label{VC} &&+G\cos^2\theta\left(f'^2_2+\frac{2\beta\omega_2}{\alpha^2+\beta^2}f'_2+\frac{\omega_2^2}{\alpha^2+\beta^2}\right)
=\frac{\kappa^2}{\alpha^2+\beta^2},
\\ \nn && \theta'^2+G\sin^2\theta\left(f'^2_1+\frac{\omega_1}{\beta}f'_1\right)
+G\cos^2\theta\left(f'^2_2+\frac{\omega_2}{\beta}f'_2\right)
 =0
.\eea

Replacing (\ref{fjs}) in the constraints (\ref{VC}) one finds the
first integral $\theta'$ of the equation of motion for $\theta$ and
a relation among the parameters \bea\label{00r}
&&\theta'^2=\frac{1}{(\alpha^2-\beta^2)^2}
\Bigg[(\alpha^2+\beta^2)\kappa^2 -\frac{C_1^2}{\sin^2\theta}
-\frac{C_2^2}{\cos^2\theta}
\\ \nn &&-\left(\alpha\omega_1+\tilde{\gamma}C_2\right)^2\sin^2\theta
-\left(\alpha\omega_2-\tilde{\gamma}C_1\right)^2\cos^2\theta \Bigg],
\\ \label{01r} && \omega_1C_1+\omega_2C_2+\beta\kappa^2=0.\eea

Now, we introduce the variable \bea\nn \chi=\cos^2\theta,\eea and
the parameters \bea\nn && v=-\frac{\beta}{\alpha},\h
u=\frac{\Omega_2}{\Omega_1},\h
W=\left(\frac{\kappa}{\Omega_1}\right)^2,\h
K=\frac{C_2}{\alpha\Omega_1},
\\ \nn &&\Omega_1=\omega_1\left(1+\tilde{\gamma}\frac{C_2}{\alpha\omega_1}\right), \h
\Omega_2=\omega_2\left(1-\tilde{\gamma}\frac{C_1}{\alpha\omega_2}\right)
.\eea By using them and (\ref{01r}), the three first integrals can
be rewritten as \bea\nn
&&f'_1=\frac{\Omega_1}{\alpha}\frac{1}{1-v^2} \left[\frac{v W-u
K}{1-\chi} -v(1-\tilde{\gamma}K)-\tilde{\gamma}u\chi\right],
\\ \label{tfid} &&f'_2=\frac{\Omega_1}{\alpha}\frac{1}{1-v^2} \left[\frac{K}{\chi}
-uv(1-\tilde{\gamma}K)-\tilde{\gamma}v^2W+\tilde{\gamma}(1-\chi)\right],
\\ \nn && \theta'
=\frac{\Omega_1}{\alpha}\frac{\sqrt{1-u^{2}}}{1-v^2}
\sqrt{\frac{(\chi_{p}-\chi)(\chi-\chi_{m})(\chi-\chi_{n})}{\chi(1-\chi)}},\eea
where \bea\nn &&\chi_p+\chi_m+\chi_n=\frac{2-(1+v^2)W-u^2}{1
-u^2},\\
\label{eqs} &&\chi_p \chi_m+\chi_p \chi_n+\chi_m
\chi_n=\frac{1-(1+v^2)W+(v W-u K)^2-K^2}{1 -u^2},\\ \nn && \chi_p
\chi_m \chi_n=- \frac{K^2}{1 -u^2}.\eea The case of dyonic
finite-size giant magnons we are interested in, corresponds to
\bea\nn 0<u<1,\h 0<v<1,\h 0<W<1,\h 0<\chi_{m}<\chi< \chi_{p}<1,\h
\chi_{n}<0.\eea

Replacing (\ref{tfid}) and (\ref{eqs}) in (\ref{r3}), one can find
the final form of the Lagrangian to be (we set $\alpha=\Omega_1=1$
for simplicity) \bea\label{Lf} &&\mathcal{L}_\gamma=
-\frac{1}{1-v^2}\left[2-(1+v^2)W-2\tilde{\gamma}K
-2\left(1-\tilde{\gamma}K-u\left(u-\tilde{\gamma}uK+\tilde{\gamma}vW\right)\right)\chi\right].\eea
After setting $\tilde{\gamma}=0$ in (\ref{Lf}) it coincides with
(\ref{Lgm}) as it should be. Now, in the $\gamma$-deformed case, the
first integral for $\chi$ reads \bea\label{fichid} \chi'=
\frac{2\sqrt{1-u^{2}}}{1-v^2}
\sqrt{(\chi_{p}-\chi)(\chi-\chi_{m})(\chi-\chi_n)}.\eea

To compute the normalized structure constant (\ref{nsc}) for the
case of two finite-size dyonic giant magnons living on the
$\gamma$-deformed three-sphere $S_{\gamma}^3$, we have to modify
(\ref{c3d}) by using (\ref{Lf}) \bea\label{c3dg} \mathcal{C}_3^q\to
\mathcal{C}_{3\gamma}^{q}=
c_{\Delta_q}\int_{-\infty}^{\infty}\frac{d\tau_e}{\cosh^{\Delta_q}(\sqrt{W}\tau_e)}
\int_{-L}^{L}d\sigma\left(\mathcal{L}_\gamma\right)^{q}.\eea
Replacing the integration over $\sigma$ according to (\ref{int}) and
using (\ref{fichid}), one finds \bea\label{c3dgf}
&&\mathcal{C}_{3\gamma}^{q}= c_{\Delta_q}\pi^{3/2}
\frac{\Gamma\left(\frac{\Delta_q}{2}\right)}{\Gamma\left(\frac{\Delta_q+1}{2}\right)}
\frac{(-2A)^q}{(1-v^2)^{q-1}\sqrt{(1-u^2)W(\chi_p-\chi_n)}}
\\ \nn &&\sum_{k=0}^{q}\frac{q!}{k!(q-k)!}\left(-\frac{B}{A}\right)^{k}\chi_p^{k}\
F_1\left(\frac{1}{2},\frac{1}{2},-k;1;1-\epsilon,1-\frac{\chi_m}{\chi_p}\right),\eea
where \bea\label{deg} &&A=1-\frac{1}{2}(1+v^2)W-\tilde{\gamma}K ,\h
B=1-\tilde{\gamma}K -u\left[u-\tilde{\gamma}(K u-v W)\right],\\
\nn &&\epsilon=\frac{\chi_{m}-\chi_{n}}{\chi_{p}-\chi_{n}}.\eea
Here, $F_1(a,b_1,b_2;c;z_1,z_2)$ is one of the hypergeometric
functions of two variables ($Appell F_1$). In writing (\ref{c3dgf}),
we used the following property of $F_1(a,b_1,b_2;c;z_1,z_2)$
\cite{PBM-III} \bea\nn F_1(a,b_1,b_2;c;z_1,z_2)=
(1-z_1)^{-b_1}(1-z_2)^{-b_2}F_1\left(c-a,b_1,b_2;c;\frac{z_1}{z_1-1},\frac{z_2}{z_2-1}\right).\eea
Then, the arguments of $F_1$, $(1-\epsilon,1-\chi_m/\chi_p)$ belong
to the interval (0,1). This representation gives the possibility to
use the defining series for $F_1(a,b_1,b_2;c;z_1,z_2)$, \bea\nn
F_1(a,b_1,b_2;c;z_1,z_2)=\sum_{k_1=\ 0}^{\infty}\sum_{k_2=\
0}^{\infty}\frac{(a)_{k_1+k_2}(b_1)_{k_1}(b_2)_{k_2}}{(c)_{k_1+k_2}}
\frac{z_1^{k_1}z_2^{k_2}}{k_1 !\ k_2 !},\h \vert z_{1}\vert <1,\h
\vert z_{2}\vert <1 ,\eea in order to consider the limits
$\epsilon\to 0$, $\chi_m/\chi_p\to 0$ , or both. The small
$\epsilon$ limit corresponds to taking into account the {\it
leading} finite-size effect, while $\epsilon=0$, $\chi_m=0$,
$\chi_n=0$, $K=0$, $W=1$, describes the infinite-size case.

Now, let us write down what the general formula (\ref{c3dgf}) for
the normalized structure constant in the $\gamma$-deformed case
gives for the first two string levels.

$q=1$ (level $n=0$): \bea\nn
&&\mathcal{C}_{3\gamma}^1=2c_{\Delta_1}\pi^{3/2}
\frac{\Gamma\left(\frac{\Delta_1}{2}\right)}{\Gamma\left(\frac{\Delta_1+1}{2}\right)}
\frac{1-\frac{1}{2}(1+v^2)W-\tilde{\gamma}K}
{\sqrt{(1-u^2)W(\chi_p-\chi_n)}}
\\ \nn &&\Bigg[\frac{1-u^2-\tilde{\gamma}\left(uvW+(1-u^2)K\right)}{1-\frac{1}{2}(1+v^2)W-\tilde{\gamma}K}
\chi_p \ F_1\left(1/2,1/2,-1;1;1-\epsilon,1-\chi_m/\chi_p\right)
\\ \nn
&&-\frac{2}{\pi}\mathbf{K}\left(1-\epsilon\right)\Bigg].\eea

$q=2$ (level $n=1$): \bea\nn
&&\mathcal{C}_{3\gamma}^2=4c_{\Delta_2}\pi^{3/2}
\frac{\Gamma\left(\frac{\Delta_2}{2}\right)}{\Gamma\left(\frac{\Delta_2+1}{2}\right)}
\frac{\left(1-\frac{1}{2}(1+v^2)W-\tilde{\gamma}K\right)^2}
{(1-v^2)\sqrt{(1-u^2)W(\chi_p-\chi_n)}}
\\ \nn &&\Bigg[\frac{2}{\pi}\mathbf{K}\left(1-\epsilon\right) -
2\frac{1-u^2-\tilde{\gamma}\left(uvW+(1-u^2)K\right)}{1-\frac{1}{2}(1+v^2)W-\tilde{\gamma}K}
\chi_p \ F_1\left(1/2,1/2,-1;1;1-\epsilon,1-\chi_m/\chi_p\right)
\\ \nn
&&+\left(\frac{1-u^2-\tilde{\gamma}\left(uvW+(1-u^2)K\right)}{1-\frac{1}{2}(1+v^2)W-\tilde{\gamma}K}\right)^2\chi_p^2
\ F_1\left(1/2,1/2,-2;1;1-\epsilon,1-\chi_m/\chi_p\right)\Bigg].\eea

Let us point out that in the $\gamma$-deformed case, we can not
obtain the reduction of (\ref{c3dgf}) to the giant magnons with one
nonzero angular momentum just by setting $u=0$. It follows from the
observation that now the smallest consistent reduction is to the
$R_t\times S_{\gamma}^3$ subspace of $AdS_5\times S_{\gamma}^5$ due
to the {\it twisted} boundary conditions \cite{BF08}. This becomes
clear if one looks at the explicit expression for the second angular
momentum \bea\nn \mathcal{J}_2\equiv\frac{2\pi
J_2}{\sqrt{\lambda}}=\frac{2}{\sqrt{1-u^2}} \left[\frac{u\chi_n-v
K}{\sqrt{\chi_{p}-\chi_{n}}} \mathbf{K}(1-\epsilon)
+u\sqrt{\chi_{p}-\chi_{n}} \mathbf{E}(1-\epsilon)\right].\eea

\setcounter{equation}{0}
\section{Concluding Remarks}
Recently, some progress have been made in computing the {\it
finite-size} effects on the three-point correlation functions in
$AdS/CFT$ context, in the framework of the semiclassical
approximation \cite{AB1105,Lee:2011,AB11062,PLB1107}. This was done
for the case when the "heavy" string states are finite-size giant
magnons, carrying one or two angular momenta, while the "light"
state was taken to be represented by the dilaton operator. The case
of "light" primary scalar operator was also investigated in
\cite{PLB1107}.

Here, we considered the case, when the "light" states are taken to
be singlet scalar operators on {\it higher} string levels
\cite{rt10}.  We first considered the case of string theory on
$AdS_5\times S^5$ dual to $\mathcal{N} = 4$ super Yang-Mills. Then
we extended the obtained results to the $\gamma$-deformed
$AdS_5\times S^5_\gamma$ background, corresponding to $\mathcal{N} =
1$ super Yang-Mills theory, arising as marginal deformation of
$\mathcal{N} = 4$ super Yang-Mills. We hope that our contribution
will help for better understanding of the $AdS/CFT$ duality at the
level of holographic correlation functions.

There are other explicitly known vertexes (see \cite{rt10} and
references therein), so it will be an interesting task to consider
the finite-size effects in these cases also.

\section*{Acknowledgements}
This work was supported in part by DO 02-257 grant.

\end{document}